\documentclass[aps,prl,reprint,amsmath,amssymb,longbibliography,nofootinbib,floatfix,superscriptaddress]{revtex4-1}
\usepackage{graphicx}
\usepackage{physics}
\usepackage{xcolor}
\usepackage{comment}
\usepackage{makecell}
\usepackage{hyperref}
\usepackage{bm}
\usepackage{letltxmacro}

\LetLtxMacro{\originaleqref}{\eqref}

\renewcommand{\Re}{\real}
\renewcommand{\Im}{\imaginary}
\providecommand{\ii}{\text{i}}
\providecommand{\ee}{\text{e}}
\newcommand{\unit}[1]{\mathbf{\hat{#1}}}
\newcommand{\Bisp}{\vec{\boldsymbol{\Psi}}}
\newcommand{\bisp}{\vec{\boldsymbol{\psi}}}

\newcommand{\Arg}{\text{Arg}}

\newcommand{\red}[1]{{\color{black}{#1}}}
\newcommand{\blue}[1]{{\color{black}{#1}}}

\begin{document}

\title{Electric-Magnetic Geometric Phase}
\author{Alex J. Vernon}
\email{alex.vernon@dipc.org}
\affiliation{Donostia International Physics Center (DIPC), Donostia-San Sebasti\'an 20018, Spain}

\author{Konstantin Y. Bliokh}
\email{konstantin.bliokh@dipc.org}
\affiliation{Donostia International Physics Center (DIPC), Donostia-San Sebasti\'an 20018, Spain}
\affiliation{IKERBASQUE, Basque Foundation for Science, Bilbao 48009, Spain}
\affiliation{Centre of Excellence ENSEMBLE3 Sp.~z o.o., 01-919 Warsaw, Poland}

\begin{abstract}
Geometric phases play an enormous role in optics and are generally associated with the evolution of light's polarization state on the Poincar\'{e} sphere, or its spin on the sphere of spin directions.
Here we put forward a new kind of optical geometric phase that appears exclusively in nonparaxial light, resulting from cyclic changes to the relative amplitude and phase between the electric and magnetic fields.
This phase is naturally represented on a recently introduced `electric-magnetic' sphere.
\end{abstract}

\maketitle

{\it Introduction.---}That the coincident initial and final states of a wave system, evolved cyclically through parameter space, can be distinguished by a geometric phase is a key concept in modern physics \cite{Berry1984, Wilczek1989_book, Zwanziger1990ARPC, Xiao2010RMP, Baggio2017JHEP, Cohen2019}.
It is difficult to overstate the importance of geometric phases in optics \cite{Vinitskii1990PU, Bhandari1997PR, Cisowski2022RMP}, underpinning the function of metasurfaces and spin-orbit coupling phenomena \cite{Hasman2005PO, Marrucci2011JO, Bliokh2015NP, Jisha2021}.

There are two basic types of optical geometric phases: (i) the Pancharatnam-Berry (PB) phase \cite{Pancharatnam1956, Berry1987, Bhandari1988PRL}  emerges when the polarization of a paraxial beam of monochromatic light is changed cyclically by propagating through a suitable series of waveplates and/or polarizers; and (ii) the spin-redirection geometric phase \cite{Rytov1938, Vladimirskiy1941, Ross1984, Tomita1986} which is accumulated during cyclic changes of the propagation direction of circularly polarized light.
These two phases are represented by closed loops and solid angles on the {\it Poincar\'{e}} and {\it spin-direction} spheres, respectively. 
Once both the 3D orientation and ellipticity of a polarization ellipse can vary, a more general approach involving a combined {\it Majorana-type} sphere \cite{Hannay1998, Bliokh2019} is needed.
\blue{Furthermore, polarization-independent optical geometric phases can appear during (i) a cyclic evolution of the modal profile (transverse phase-intensity distribution) of a paraxial beam \cite{vanEnk1993, Galvez2003, Calvo2005}, which is naturally represented on the modal analogue of the Poincar\'{e} sphere \cite{Padgett1999, Calvo2005, Dennis2017}; and (ii) variations of the propagation direction of vortex beams carrying orbital angular momentum \cite{Bliokh2006PRL, Alexeyev2006JOA, Kataevskaya1995QE, Wang2018SA}.}

Each of the above kinds of optical geometric phase are described by evolutions of the polarization and intensity-phase distribution of the wave's {\it electric} field.
Yet, just as the electric field's polarization ellipse is transformed, so too is that of the {\it magnetic} field---and in a different fashion in general. 
Moreover, Maxwell's equations permit, in nonparaxial light, the electric and magnetic fields a surprising amount of freedom: both the share of energy between the two fields and their relative phase can change.
This is analogous to how the two components of a paraxial beam's polarization (Jones vector) change in amplitude and relative phase in the PB phase problem. 

In this article, we show that such a relative evolution of the electric and magnetic fields in a nonparaxial optical field gives rise to a new ``{\it electric-magnetic}" (EM) geometric phase, described within a generalized theory of optical geometric phase for the six-component vector that combines the two fields.
Most startlingly, we show that the EM geometric phase can even present itself when the polarization states of the electric and magnetic fields \textit{do not change}, providing a simple example of this scenario: a standing wave modulated cyclically in time.
In a second example, we show the remarkable emergence of the EM geometric phase in the longitudinal field components in a focused beam whose spatial structure evolves cyclically---a link between the modal geometric phase for paraxial beams and the strong-focusing regime.
\blue{Notably, the geometric phase described in our work is of an Aharonov-Anandan type \cite{Aharonov1987}, which requires cyclic evolution of the state vector but not adiabaticity.}   

{\it General theory.---}An electromagnetic field as a function of some parameters $\bm{\rho}$ can be expressed with a single vector in the $\mathbb{C}^2\otimes\mathbb{C}^3\otimes\mathbb{L}^2$ space \cite{BialynickiBirula1996, Berry2009, Bliokh2014, Golat2025}, 
\begin{equation}
\label{bispinor}
\begin{split}    \Bisp(\bm{\rho})&=\frac12\!\begin{pmatrix}\sqrt{\varepsilon_0}\mathbf{E}(\bm{\rho})\\\sqrt{\mu_0}\mathbf{H}(\bm{\rho})\end{pmatrix}=\sqrt{W(\bm{\rho})}\ee^{\ii\alpha(\bm{\rho})}\bisp(\bm{\rho})\,,
\end{split}
\end{equation}
where $\mathbf{E}(\bm{\rho})$ and $\mathbf{H}(\bm{\rho})$ are the complex electric and magnetic fields,  $W=\Bisp^\dagger\cdot\Bisp=\left(\varepsilon_0\mathbf{E}^*\cdot\mathbf{E}+\mu_0\mathbf{H}^*\cdot\mathbf{H}\right)/4$
is the electromagnetic energy density, $\bisp$ is a normalized vector ($\bisp^\dagger\cdot\bisp=1$), and the overall phase $\alpha$ is to be chosen.
Here and in what follows, $\dagger$ and $\intercal$ superscripts mean conjugate transpose and transpose, respectively, bold characters like $\mathbf{E}$ are $\mathbb{C}^3$ vectors while those with arrow accents like $\Bisp$ are $\mathbb{C}^2\otimes\mathbb{C}^3$ vectors, and explicit expression of the $\bm{\rho}$-dependence of quantities in Eq.~\eqref{bispinor} is to be avoided except for useful emphasis.
All phase angles in this work are defined within the interval $[0,2\pi)$---any expression for a phase angle is assumed to be taken modulo $2\pi$.

\begin{figure}[t]
\centering
\includegraphics[width=\columnwidth]{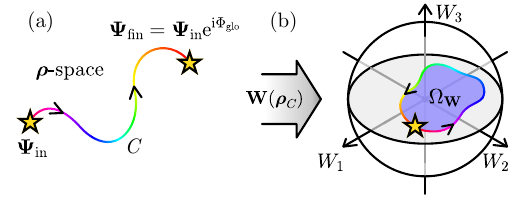}
\caption{General geometric phase scenario. (a) The field $\Bisp$, Eq.~\eqref{bispinor} evolves over a curve $C$ in the parameter $\bm{\rho}$ space.
At the two ends of $C$ the field differs only by the phase  $\Phi_\text{glo}=\Phi_\text{loc}+\Phi_\text{g}$, Eqs.~\eqref{Pan_phase}, \eqref{local_phase}, and \eqref{geom_phase_general}.
(b) If the electric and magnetic polarization states do not vary along $C$, then $\Phi_\text{g}=-\Omega_\mathbf{W}/2$ is the EM geometric phase, proportional to the solid angle enclosed by the evolution loop on a Poincar\'e-like EM sphere \cite{Golat2025}, Eqs.~\eqref{Wvec} and \eqref{solid_angle}.}
\label{fig1}
\end{figure}

Suppose that there is an open curve \blue{$C$ in parameter $\bm{\rho}$ space,} along which $\Bisp$ is varied between an initial value $\Bisp_\text{in}=\Bisp(\bm{\rho}_\text{in})$ and a final value $\Bisp_\text{fin}=\Bisp(\bm{\rho}_\text{fin})$, distinguishable only by a {\it global} phase: $\Bisp_\text{fin}=\Bisp_\text{in}\exp(\ii\Phi_\text{glo})$ \cite{Aharonov1987}, see Fig.~\ref{fig1}(a).
It is straightforward to see that this phase factor \blue{is consistent with} Pancharatnam's definition \cite{Pancharatnam1956,Berry1987} of the phase difference between two non-orthogonal field vectors:
\begin{equation}\label{Pan_phase}
\begin{split}
\Phi_\text{glo}&=\Arg{\left(\Bisp^\dagger_\text{in}\cdot\Bisp_\text{fin}\right)}=
\Delta\alpha+\Arg{\left(\bisp^\dagger_\text{in}\cdot\bisp_\text{fin}\right)}\,,
\end{split}
\end{equation}
where $\Delta\alpha=\alpha(\bm{\rho}_\text{fin})-\alpha(\bm{\rho}_\text{in})$.
Since the field evolves continuously over $C$ it is also possible to record \textit{local} phase accumulation.
An incremental change in phase between neighbouring points in $\bm{\rho}$-space is, using Pancharatnam's definition \eqref{Pan_phase},
\begin{equation}\label{local_k}
\begin{split}
d\Phi_\text{loc}\!=\Arg\!\left[\Bisp^\dagger\cdot ( \Bisp+d\Bisp )\right]\!=\!\underbrace{\frac{1}{W}\Im\!\left[\Bisp^\dagger\cdot(\bm{\nabla}_{\bm{\rho}})\Bisp\right]}_{\mathbf{k}_{\bm{\rho}}}\cdot\, d\bm{\rho}\,.
\end{split}
\end{equation}
Interpreting $\mathbf{k}_{\bm{\rho}}$ as a ``local wavevector'' over $\bm{\rho}$ \cite{Bliokh2019, Berry2019, Berry2009}, 
we might initially expect that its integral along $C$ would recover $\Phi_\text{glo}$, Eq.~\eqref{Pan_phase}.
But when substituting Eq.~\eqref{bispinor}, its integral,
\begin{equation}\label{local_phase}
\Phi_\text{loc}=\Delta \alpha-\ii\int_C\bisp^\dagger\cdot(\bm{\nabla}_{\bm{\rho}})\bisp\cdot d\bm{\rho}\,,
\end{equation}
is \textit{not} generally equal to $\Phi_\text{glo}$.
Indeed, the difference between Eqs.~\eqref{Pan_phase} and \eqref{local_phase} is a \textit{geometric phase}: $\Phi_\text{g}=\Phi_\text{glo}-\Phi_\text{loc}$.
Geometric phase is what effects ``global change without local change" \cite{Berry2024}.
The sum effect of local change, $\Phi_\text{loc}$, can be identified with \textit{dynamical phase}.

We have yet to exercise our freedom to select the phase angle $\alpha$.
While the difference $\Phi_\text{g}=\Phi_\text{glo}-\Phi_\text{loc}$ is invariant \blue{with respect} to changes of $\alpha$ (i.e., $U(1)$ gauge transformations), its choice can be leveraged to obtain a more satisfying expression in terms of the normalized field $\bisp$.
\blue{Since $\alpha$ is an overall phase ascribed to all six field components, it is natural to define it via an associated complex scalar quadratic field $\Psi = \Bisp^\intercal\cdot\Bisp$:  
$\alpha=\Arg(\Psi)/2$ \cite{Bliokh2019, Berry2019, BerryDennis2001}.
Then it follows from Eq.~\eqref{bispinor} that $\Arg(\bisp^\intercal\cdot\bisp)=0$, and subsequently in Eq.~\eqref{Pan_phase}, ${\Arg}{(\bisp^\dagger_\text{in}\cdot\bisp_\text{fin})}=N\pi$ (see SI Eqs.~(S4)-(S7)).
Here $N=0$ or $1$ is a $\mathbb{Z}_2$ topological index which encodes the even or odd number of half-turns made by principal semiaxes the `six-dimensional EM polarization ellipse' \cite{Vernon2025_1} of $\bisp$ over $C$.
For a closed contour $C$, $N=1$ corresponds to {\it polarization M\"{o}bius strips} \cite{Freund2010OC, Bauer2015S, Bliokh2019, Bliokh2021POF, Muelas2022PRL}.}

Subtracting now Eq.~\eqref{local_phase} from Eq.~\eqref{Pan_phase}, we obtain:
\begin{equation}
\label{geom_phase_general}    \Phi_\text{g}=\ii\int_C\bisp^\dagger\cdot(\bm{\nabla}_{\bm{\rho}})\bisp\cdot d\bm{\rho}+N\pi\,.
\end{equation}
Equation (\ref{geom_phase_general}) is the central general expression in our work, describing the optical geometric phase that incorporates both electric and magnetic fields, valid for both paraxial and nonparaxial waves, and eluded to in  \cite{Bliokh2019, Berry2019}.
\blue{Note that our terminology differs from previous studies \cite{Bliokh2019, Berry2019} that refer to the integrated local wavevector as \textit{total} phase, while $\Delta\alpha$ is associated with the {\it dynamical} phase, so that the geometric phase is given in a form similar to the integral term in Eq.~\eqref{geom_phase_general} with opposite sign.} 

A natural next step is to attempt to divide $\Phi_\text{g}$ into electric and magnetic contributions as is customarily done for many quantities in electromagnetism \cite{Berry2009, Bliokh2013, Bliokh2014, Berry2019, Cameron2012, BialynickiBirula1996}.
We express the normalized state vector $\bisp$ as
\begin{align}\label{factorisation}
    \bisp=\begin{pmatrix}
        \cos{\frac{\theta}{2}}\ee^{\ii\chi_\text{e}}\unit{e}\\\sin{\frac{\theta}{2}}\ee^{\ii\chi_\text{m}}\unit{h}
    \end{pmatrix},
\end{align}
with $\theta=2\arctan{(\sqrt{\mu_0}|\mathbf{H}|/\sqrt{\varepsilon_0}|\mathbf{E}|)}$ \blue{characterizing the relative amplitude of the electric and magnetic components}, and the complex unit polarization vectors  $\unit{e}=\exp[-\ii(\alpha+\chi_\text{e})]\mathbf{E}/|\mathbf{E}|$ and $\unit{h}=\exp[-\ii(\alpha+\chi_\text{m})]\mathbf{H}/|\mathbf{H}|$ whose phases $\chi_\text{e}$ and $\chi_\text{m}$ are added to $\alpha$ such that $\Arg{(\unit{e}\cdot\unit{e})}=\Arg{(\unit{h}\cdot\unit{h})}=0$.
This fixes the real and imaginary parts of $\unit{e}$ ($\unit{h}$) at right angles, aligned to the semi-major and semi-minor axes of the electric (magnetic) polarization ellipse, respectively \cite{BerryDennis2001, Bliokh2019}.
Surprisingly, substituting Eq.~\eqref{factorisation} into Eq.~\eqref{geom_phase_general} yields two kinds of geometric phase terms:
\begin{equation}
\label{phig_decomp}
\Phi_\text{g}=\Phi_\text{gI}+\Phi_\text{gII}+N\pi\,,
\end{equation}
where
\begin{equation}\label{phi_gI}
\begin{split}
\Phi_\text{gI}
=\ii\int_C\!\left[\cos^2\!{\frac{\theta}{2}}\unit{e}^*\cdot(\bm{\nabla}_{\bm{\rho}})\unit{e}
+\sin^2\!{\frac{\theta}{2}}\unit{h}^*\cdot(\bm{\nabla}_{\bm{\rho}})\unit{h}\right]\!\cdot d\bm{\rho}\,,\\
\end{split}
\end{equation}
\blue{is a weighted sum of the electric and magnetic terms, each similar in structure to the integral term in Eq.~\eqref{geom_phase_general}, which describes geometric phase arising from variations of the electric and magnetic polarization ellipses along $C$.
This type of geometric phase has been described in \cite{Bliokh2019} (for closed-contour evolutions in real space); it accounts for both the PB phase in 2D paraxial waves, where $\theta=\pi/2$ and $\Phi_\text{gI}+N\pi=\Phi_\text{PB}$, and the spin-redirection phase for 3D evolutions of polarization.}
Having fixed $\Arg{(\unit{e}\cdot\unit{e})}=\Arg{(\unit{h}\cdot\unit{h})}=0$ means that $\Phi_\text{gI} = 0$ if the electric and magnetic polarization states are uniform over $C$.

Our main result is what we term the {\it EM geometric phase}:
\begin{equation}\label{em_phase}
\begin{split}
\Phi_\text{gII}
=-\int_C\left[\cos^2{\frac{\theta}{2}}\,d\chi_\text{e}+\sin^2{\frac{\theta}{2}}\,d\chi_\text{m}\right]\,.
\end{split}
\end{equation}
It originates from the incorporation of the $\mathbb{C}^2$ EM space in $\Bisp$ and arises \textit{exclusively in non-paraxial light}, i.e., $\Phi_\text{gII}=0$ for paraxial waves (see SI).
Perhaps the most striking feature of $\Phi_\text{gII}$ is its independence from $\mathbb{C}^3$ polarization---in nonparaxial light, a geometric phase may emerge even when the electric and magnetic polarization states are constant, i.e., neither $\unit{e}$ nor $\unit{h}$ evolve over $C$.
When, moreover, the electric and magnetic fields share the same polarization state throughout, $\unit{e}\equiv \unit{h}$, the phase $\Phi_\text{gII}$ may be related to the trajectory of the tip of a Stokes-like \red{vector} $\mathbf{W}=\Bisp^\dagger\cdot(\hat{\boldsymbol{\sigma}})\Bisp / W$ \cite{Bliokh2014, Golat2025}:
\begin{equation}\label{Wvec}
\begin{split}    
    \mathbf{W}
    =\frac{1}{4W}\!\begin{pmatrix}
        2\Re\{\mathbf{E}^*\cdot\mathbf{H}\}/\omega c\\
        2\Im\{\mathbf{E}^*\cdot\mathbf{H}\}/\omega c\\
        \varepsilon_0|\mathbf{E}|^2-\mu_0|\mathbf{H}|^2
    \end{pmatrix}=\begin{pmatrix}
        \sin{\theta}\cos{\phi}\\\sin{\theta}\sin{\phi}\\\cos\theta\end{pmatrix},  
\end{split}
\end{equation}
where $\hat{\boldsymbol{\sigma}}=(\hat{\sigma}_1,\hat{\sigma}_2,\hat{\sigma}_3)$ is the vector of Pauli matrices and $\phi=\Arg(\mathbf{E}^*\cdot\mathbf{H})=\chi_\text{m}-\chi_\text{e}$.
In this case, 
it can be shown that (see SI)
\begin{equation}
\label{solid_angle}
\begin{split}
\Phi_\text{g} = \Phi_\text{gII} +N\pi =-\frac12\int_C(1-\cos\theta)d\phi=-\frac12\Omega_\mathbf{W}\,,
\end{split}
\end{equation}
where $\Omega_\mathbf{W}$ is the solid angle swept out by $\mathbf{W}$, see Fig.~\ref{fig1}(b).
The vector $\mathbf{W}$ can be interpreted as a Stokes vector for an {\it EM ellipse} \cite{Vernon2025_1}, represented on a Poincar\'e-like {\it EM sphere}, originally proposed in \cite{Golat2025}
(the sphere in Fig.~\ref{fig1}(b) is rotated compared to that in \cite{Golat2025} because of our use of an electric-magnetic basis for $\Bisp$).
In this representation, the geometric phase \eqref{solid_angle} takes on the standard PB-like form, associated with the SU(2) evolution of the $\mathbb{C}^2$ EM vector.

So far, we have worked tacitly within an EM basis for $\Bisp$.
Yet $\Bisp$ can be expressed in a general basis \cite{Golat2025}, $\Bisp_\text{uv}=(\mathbf{F}_\text{u},\mathbf{F}_\text{v})/2$, related to $\Bisp$ via a SU(2) gauge transformation mixing $\mathbf{E}$ and $\mathbf{H}$ (i.e., rotating the axes of the EM ${\bf W}$-sphere).
This generally results in redistribution of $\Phi_\text{g}$ between the contributing terms $\Phi_\text{gI}$ and $\Phi_\text{gII}$, expressed via new fields $\mathbf{F}_\text{u}$ and $\mathbf{F}_\text{v}$, but the total geometric phase itself \eqref{geom_phase_general} and \eqref{phig_decomp} remains \textit{gauge-invariant}.
\red{When $\unit{e}\equiv\unit{h}$, meanwhile, the EM geometric phase $\Phi_\text{gII}$ also becomes gauge-invariant.}

We next propose two scenarios in which the EM geometric phase $\Phi_\text{gII}$ can be observed.

\begin{figure}[t]
    \centering
    \includegraphics[width=\columnwidth]{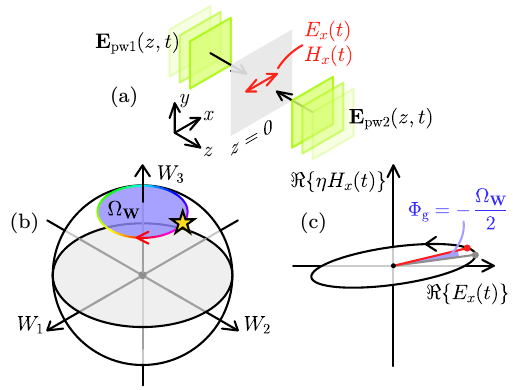}
    \caption{EM geometric phase emerging in a time-modulated standing wave.
    (a) Diagram of scenario, designed so $\mathbf{E}$ and $\mathbf{H}$ are $\unit{x}$-polarized at $z=0$, Eq.~\eqref{standing_wave}.
    (b) Trajectory of $\mathbf{W}$ (the star denotes $\mathbf{W}(0)=\mathbf{W}(T)$), corresponding to Eq.~\eqref{solution} with $\theta_0=\pi/6$ and $\chi_0=2\pi/3$ and enclosing solid angle $\Omega_\mathbf{W}\simeq -0.27$ sr.
    (c) Initial/final EM ellipse \cite{Vernon2025_1} at $z=0$.
    The red line is the instantaneous vector $(\Re\{E_x(T)\},\Re\{\eta H_x(T)\})$ of the standing wave having acquired a phase of $-\omega_0T+\Phi_\text{g}$.
    The grey line is that same vector for a reference standing wave that gains the phase $-\omega_0T$.
    }\label{fig2}
\end{figure}

{\it EM geometric phase in a standing wave.---}Whereas the usual PB and spin-redirection geometric phases can be measured with a standard interferometry setup for paraxial light, the EM geometric phase can only manifest in nonparaxial light and would require measurement of interference at a localized point in space. 
Here we consider the simplest example of a nonparaxial field: a standing wave formed by two counter-propagating plane waves along the $z$ axis, Fig.~\ref{fig2}(a).
We choose time as the parameter that drives the evolution: $\boldsymbol{\rho} = t$. 

The electric and magnetic fields of such a time-varying standing wave at $z=0$ can be written as:
\begin{equation}
\label{standing_wave_general}
\begin{split}
\mathbf{E}(t)&=\ee^{-\ii\omega_0 t}\left[\left(A_x(t)+B_x(t)\right)\unit{x}+\left(A_y(t)+B_y(t)\right)\unit{y}\right],\\
\mathbf{H}(t)&=\frac{\ee^{-\ii\omega_0 t}}{\eta}\left[\left(B_y(t)-A_y(t)\right)\unit{x}+\left(A_x(t)-B_x(t)\right)\unit{y}\right],
\end{split}
\end{equation}
where $\omega_0$ is the central frequency of the waves, $A_i(t)$ ($B_i(t)$) are the complex electric field amplitudes of the plane wave propagating in the $+z$ ($-z$) direction, which are modulated in time, 
and $\eta=\sqrt{\mu_0/\varepsilon_0}$.
By enforcing $A_x(t)=B_x(t)$ and $A_y(t)=-B_y(t)$, both electric and magnetic fields in the standing wave become purely $\unit{x}$-polarized at $z=0$ for any value of $t$.
Then the field \eqref{bispinor} becomes
\begin{equation}\label{standing_wave}
\begin{split}
    \Bisp(t)=\sqrt{W}\ee^{-\ii\omega_0 t}\begin{pmatrix}
        \cos{\frac{\theta(t)}{2}}\ee^{\ii\chi_x(t)}\unit{x}\\-\sin{\frac{\theta(t)}{2}}\ee^{\ii\chi_y(t)}\unit{x}
    \end{pmatrix}.
\end{split}
\end{equation}
%
where $\theta(t)=2\arctan(|A_y(t)|/|A_x(t)|)$ and $\chi_{x,y}(t)=\Arg(A_{x,y}(t))$.
Since modulation is carried out over time, the ``local wavevector'' \eqref{local_k} with $\nabla_{\bm{\rho}}=\partial/\partial t$ is related to the instantaneous angular frequency of the field: $k_t = \partial\Phi_\text{loc}/\partial t  = - \omega(t)$.

The geometric phase can be measured against an unmodulated reference wave that  accrues a dynamical phase of $-\omega_0T$ over some interval $t\in[0,T]$.
To ensure that the modulated wave has the same dynamical phase, we require $\partial \Phi_\text{loc}/\partial t=-\omega_0$, for which one simple solution is
\begin{equation}
\label{solution}
\chi_x(t)=\frac{2\pi t}{T}\sin^2{\frac{\theta_0}{2}},\quad\chi_y(t)=-\frac{2\pi t}{T}\cos^2{\frac{\theta_0}{2}}+\chi_0\,,
\end{equation}
with $\theta(t)=\theta_0$ kept constant, and where $\chi_0$ determines the initial phase offset between the electric and magnetic fields.
Substituting Eq.~\eqref{solution} into Eq.~\eqref{standing_wave} and calculating the global phase \eqref{Pan_phase} of the field between $t=0$ and $t=T$, we derive:
\begin{equation}
\label{standing_wave_global}
\begin{split}
\Arg\!\left(\Bisp^\dagger(0)\cdot\Bisp(T)\right)=\underbrace{-\omega_0T}_{\Phi_\text{loc}}+\underbrace{(1-\cos{\theta_0})\pi}_{\Phi_\text{g}}\,.
\end{split}
\end{equation}
This clearly contains a dynamical phase $\Phi_\text{loc}$ and a geometric phase $\Phi_\text{g}$ that is independent of the modulation time interval $T$ and corresponds to the solid angle enclosed by the evolution contour on the EM ${\bf W}$-sphere, Fig.~\ref{fig2}(b).
It is straightforward also to obtain $\Phi_\text{g}$ using the integral \eqref{solid_angle} over $t\in [0,T]$ with $\theta=\theta_0$ and $d\phi=d\chi_\text{m}-d\chi_\text{e} \equiv d\chi_{y}-d\chi_{x}
=-(2\pi/T)\,dt$.
In Fig.~\ref{fig2}(c), $\Phi_\text{g}$ is visualized as a phase gained by an EM ellipse \cite{Vernon2025_1}, whose shape before and after modulation is identical, against a reference standing wave.

We emphasise that the geometric phase in Eq.~\eqref{standing_wave_global} arises despite both electric and magnetic fields having constant $\unit{x}$ polarization.
\blue{Moreover, Eq.~\eqref{standing_wave_global} is exact, as in the Aharonov-Anandan approach \cite{Aharonov1987}, and does not require time modulation to be slow. This is in contrast to the adiabatic geometric phase \cite{Berry1984}, which is just the zero-order term in an infinite series over $1/T$ \cite{Berry1987_2, Bliokh2002}.}

\begin{figure}[t]
\centering
\includegraphics[width=\columnwidth]{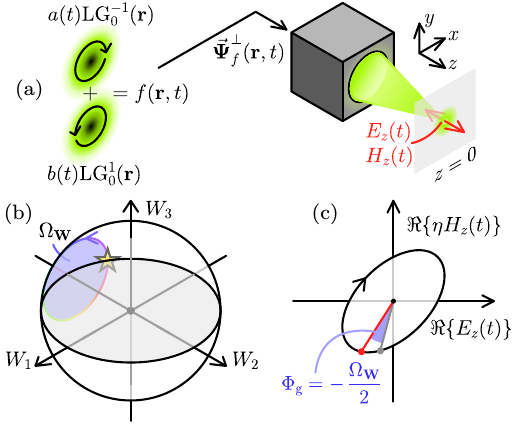}
\caption{EM geometric phase arising in the longitudinal components of a time-evolving focused beam carrying a first-order modal structure.
(a) Schematic, where a linearly polarized beam whose mode profile $f$ is modulated over $t$, Eq.~\eqref{bisp_f}, is focused to produce longitudinal components \eqref{z_comp} in the focal center $\mathbf{r=0}$.
(b) Closed trajectory of $\mathbf{W}$ corresponding to Eqs.~\eqref{z_comp} and \eqref{ab_focused} with $\theta_0=\pi/6$ and $\chi_0=\pi/2$.
(c) Final state of the ellipse formed of the electric and magnetic z components in the beam focus, annotated similarly as in Fig.~\ref{fig2}(c).}
\label{fig3}
\end{figure}

{\it EM geometric phase in focused beams.---}Another manifestation of the EM geometric phase can be found in the longitudinal components of a strongly focused beam with mode order 1 (such as Hermite-Gauss $\text{HG}_{10}$ and $\text{HG}_{01}$ or Laguerre-Gauss $\text{LG}_{0}^{\pm 1}$ modes, and their superpositions), whose transverse spatial structure is transformed cyclically prior to focusing, Fig.~\ref{fig3}(a).
Cyclic mode transformations made to paraxial beams are well-known to produce polarization-independent geometric phases \cite{vanEnk1993, Galvez2003, Calvo2005} associated with closed paths on the modal Poincar\'e sphere \cite{Padgett1999, Calvo2005, Dennis2017}.
A focused beam of light, however, develops {\it longitudinal} electric and magnetic field components that are not accounted for by the modal Poincar\'e sphere but which can be made to evolve in time (${\bm{\rho}} = t$ as in the previous example) by changing the input paraxial beam's mode structure, and thereby producing the EM geometric phase.

A general first-order mode $f$ can be presented as a linear superposition of $\text{LG}^{\pm 1}_0$ modes.
The transverse field of a time-varying first-order beam is given by the four-vector
\begin{equation}
\label{bisp_f}
\begin{split}
\Bisp^\perp_f(\mathbf{r},t)&=\ee^{-\ii\omega_0 t}\Bisp_0^\perp\underbrace{\left[a(t)\text{LG}^{-1}_0(\mathbf{r})+b(t)\text{LG}^{1}_0(\mathbf{r})\right]}_{f(\mathbf{r},t)}\,,
\end{split}
\end{equation}
where 
$a(t)$ and $b(t)$ are complex time-varying coefficients satisfying $|a|^2+|b|^2=1$ (for example, $|a|=|b|$ corresponds to Hermite-Gauss modes),
and the vector $\Bisp_0^\perp=(\mathbf{F}_{\text{R}0}^\perp,\mathbf{F}_{\text{L}0}^\perp)/2$ determines the constant transverse polarization, here expressed in a basis of Riemann-Silberstein vectors \cite{BialynickiBirula1996, Golat2025} $\mathbf{F}_{\text{R0,L0}}^\perp=(\sqrt{\varepsilon_0}\mathbf{E}^\perp_0\pm\ii\sqrt{\mu_0}\mathbf{H}^\perp_0)/\sqrt{2}$.
Due to paraxiality, $\mathbf{F}_\text{R0,L0}^\perp=A_\text{R,L}^\perp(\unit{x}\pm\ii\unit{y})/\sqrt{2}$ are oppositely circularly polarized where $A_\text{R,L}^\perp$ are complex constants.

Once strongly focused, the longitudinal ($\unit{z}$-polarized) field, $\Bisp_f^\parallel$, strongest in the focal plane at the beam's center $\mathbf{r=0}$, can be approximated as \cite{Green2023, Forbes2025}  $\Bisp_f^\parallel\simeq \ii k^{-1}(\bm{\nabla}_\perp\cdot\Bisp_f^\perp)\,\unit{z}$, which yields (see SI):
%
\begin{equation}
\label{z_comp}
    \Bisp_f^\parallel(\mathbf{0},t)=\frac12\begin{pmatrix}F_{\text{R}z}(t)\unit{z}\\F_{\text{L}z}(t)\unit{z} 
    \end{pmatrix}=\frac{\ii B\ee^{-\ii\omega_0 t}}{k\sqrt{2}}\begin{pmatrix}
       A_\text{R}^\perp a(t)\unit{z}\\A_\text{L}^\perp b(t)\unit{z}
    \end{pmatrix}\,.
\end{equation}
where $B$ is a constant.
Remarkably, Eq.~\eqref{z_comp} shows that the longitudinal components of right- and left-handed Riemann-Silberstein vectors in the center of the beam's focus are \textit{independently} modulated by the complex functions $a(t)$ and $b(t)$ that determine the beam's paraxial spatial structure \eqref{bisp_f}.
It resembles Eq.~\eqref{standing_wave} of the standing-wave scenario, except that the entries of $\Bisp^\parallel_f$ are not electric and magnetic fields. 
Choosing
\begin{equation}
\label{ab_focused}
\begin{split}
a(t)=\cos\frac{\theta_0}{2}\ee^{\ii\chi_a}\,,\quad
b(t)=\sin\frac{\theta_0}{2}\ee^{\ii\chi_b}\,,
\end{split}
\end{equation}
where $\chi_a$ and $\chi_b$ coincide with $\chi_x$ and $\chi_y$ in Eqs.~\eqref{solution}, the EM geometric phase can be obtained over $t\in[0,T]$ from the equivalent of Eq.~\eqref{solid_angle} in the Riemann-Silberstein basis if $|A_{\text{R}}^\perp|=|A_{\text{L}}^\perp|$ (linear transverse electric polarization):
\begin{equation}
\label{geomphase_focused}
\begin{split}
\Phi_\text{g}&=-\frac12\int_0^T(1-\cos\theta_\text{RL})\, d\phi_\text{RL} =(1-\cos\theta_0)\pi,
\end{split}
\end{equation}
where $\theta_\text{RL}=\theta_0$ is the polar angle defined with respect to the $-W_2$ axis on the EM sphere, and $d\phi_\text{RL} \equiv d\chi_b - d\chi_a = -(2\pi/T) dt$ is the differential of the corresponding azimuthal angle, see Fig.~\ref{fig3}(b).
Equation~\eqref{geomphase_focused} is equivalent to what would be the modal geometric phase \cite{vanEnk1993, Galvez2003, Calvo2005, Padgett1999} in an unfocused paraxial beam, \blue{but now it is described by the evolution of the field in a {\it single point} rather than by a spatially-extended modal structure}.
Figure~\ref{fig3}(b,c) shows the ${\bf W}$-sphere representation of the EM geometric phase  \eqref{geomphase_focused}, which is similar to the standing-wave scenario, Fig.~\ref{fig2}(b,c), up to a global rotation about the $W_1$ axis because of the use of the Riemann-Silberstein basis.

{\it Conclusions.---}
We have presented a generalized theory of optical geometric phase, that incorporates both electric and magnetic fields consistently, and 
identified a new electric-magnetic (EM) geometric phase.
Together with the PB and spin-redirection phases, it completes the picture of optical geometric phase.
The EM geometric phase emerges only in nonparaxial light and results from cyclic evolutions of the relative amplitude and phase between the electric and magnetic fields. Remarkably, it can be observed in nonparaxial waves where both fields have the same, constant polarization ellipse, but only vary in relative amplitude and phase.
We have presented two examples of this scenario: a standing wave modulated in time, and the focal center of a strongly focused first-order beam, whose spatial structure is modulated in time.


Importantly, although the theoretical description of the EM geometric phase involves both electric and magnetic fields, it can be observed experimentally as the phase of the electric field. Moreover, one can analyse the wave phase using an electric-field-only formalism, as in \cite{Bliokh2019}; in this case the EM geometric phase will be subsumed into the electric field's dynamical phase analogous to Eq.~\eqref{local_phase}.

Optical geometric phases have proved invaluable for the design of advanced optical elements and metasurfaces \cite{Hasman2005PO, Marrucci2011JO, Bliokh2015NP, Jisha2021}.
Much current effort is now devoted to both time-varying materials \cite{Galiffi2022AP} and time-structured waves \cite{Yessenov2022AOP, Shen2023JO}.
The EM geometric phase could aid in the design of new generations of metamaterials. Our formalism also ties into recent studies of the topological features of the six-component electromagnetic field \eqref{bispinor} \cite{Vernon2025, Vernon2025_1}.


We finally remark that an interesting recent preprint \cite{Cheng2025} has suggested that a new ``Riemann-Silberstein geometric phase" emerges in {\it paraxial} light when working in a basis of Riemann-Silberstein-like vectors that mix {\it different} components of the $\mathbf{E}$ and $\mathbf{H}$ fields. In this case, however, the electric and magnetic polarizations are not locally independent, and their proposed geometric phase would appear to conflict with our discussion of the equivalence of the total geometric phase to the known PB phase for paraxial waves. 

\begin{acknowledgments}
This work is supported in part by the Marie Sk\l{}odowska-Curie COFUND Programme of the European Commission (project HORIZON-MSCA-2022-COFUND-101126600-SmartBRAIN3), 
ENSEMBLE3 Project (MAB/2020/14) which is carried out within the International Research Agendas Programme (IRAP) of the Foundation for Polish Science co-financed by the European Union under the European Regional Development Fund and Teaming Horizon 2020 programme (GA. No. 857543) of the European Commission, and the project of the Minister of Science and Higher Education ``Support for the activities of Centers of Excellence established in Poland under the Horizon 2020 program'' (contract MEiN/2023/DIR/3797). 
\end{acknowledgments}

\bibliography{bibliography}

@article{Bliokh2019,
   author = {K. Y. Bliokh and M. A. Alonso and M. R. Dennis},
   doi = {10.1088/1361-6633/ab4415},
   journal = {Rep. Prog. Phys.},
   pages = {122401},
   title = {{Geometric phases in 2D and 3D polarized fields: geometrical, dynamical, and topological aspects}},
   volume = {82},
   year = {2019}
}

@article{Berry1984,
	author = {Berry, M. V.},
	title = {{Quantal phase factors accompanying adiabatic changes}},
	journal = {Proc. R. Soc. Lond. A.},
	volume = {392},
	pages = {45},
	year = {1984},
	doi = {10.1098/rspa.1984.0023}
}

@article{Berry1987_2,
	author = {Berry, M. V.},
	title = {Quantum Phase Corrections from Adiabatic Iteration},
	journal = {Proc. R. Soc. Lond. A.},
	volume = {414},
	pages = {31},
	year = {1987},
	doi = {10.1098/rspa.1987.0131}
}

@article{Bliokh2002,
    author = {Bliokh, K. Y.},
    title = {{Generalization of Berry’s geometric phase, equivalence of the Hamiltonian nature, quantizability and strong stability of linear oscillatory systems, and conservation of adiabatic invariants}},
    journal = {J. Math. Phys.},
    volume = {43},
    pages = {5624},
    year = {2002},
    doi = {10.1063/1.1506954},
}

@article{Alexeyev2006JOA,
doi = {10.1088/1464-4258/8/9/008},
year = {2006},
volume = {8},
pages = {752},
author = {Alexeyev, C. N. and Yavorsky, M. A.},
title = {Topological phase evolving from the orbital angular momentum of ‘coiled’ quantum
vortices},
journal = {J. Opt. A: Pure Appl. Opt.}
}

@article{Bliokh2006PRL,
  title = {{Geometrical Optics of Beams with Vortices: Berry Phase and Orbital Angular Momentum Hall Effect}},
  author = {Bliokh, K. Y.},
  journal = {Phys. Rev. Lett.},
  volume = {97},
  pages = {043901},
  year = {2006},
  doi = {10.1103/PhysRevLett.97.043901}
}

@article{Kataevskaya1995QE,
doi = {10.1070/QE1995v025n09ABEH000504},
year = {1995},
volume = {25},
pages = {927},
author = {I. V. Kataevskaya and N. D. Kundikova},
title = {Influence of the helical shape of a fibre waveguide on the propagation of light},
journal = {Quantum Electron.},
}

@article{Wang2018SA,
author = {S. Wang  and G. Ma  and C. T. Chan },
title = {Topological transport of sound mediated by spin-redirection geometric phase},
journal = {Sci. Adv.},
volume = {4},
pages = {eaaq1475},
year = {2018},
doi = {10.1126/sciadv.aaq1475},
}

@article{Pancharatnam1956,
	author = {Pancharatnam, S.},
	title = {{Generalized theory of interference, and its applications}},
	journal = {Proc. Indian Acad. Sci.},
	volume = {44},
	pages = {247},
	year = {1956},
	doi = {10.1007/BF03046050}
}

@article{BerryDennis2001,
	Author = {M. V. Berry and M. R. Dennis},
	Journal = {Proc. Roy. Soc. Lond. A},
	Pages = {141},
	Title = {Polarization singularities in isotropic random vector waves},
	Volume = {457},
	Year = {2001},
	doi = {10.1098/rspa.2000.0660}
}

@article{Berry1987,
   author = {M. V. Berry},
   doi = {10.1080/09500348714551321},
   issn = {0950-0340},
   issue = {11},
   journal = {J. Mod. Opt.},
   month = {11},
   pages = {1401},
   title = {{The Adiabatic Phase and Pancharatnam's Phase for Polarized Light}},
   volume = {34},
   year = {1987}
}

@article{Cohen2019,
   author = {E. Cohen and H. Larocque and F. Bouchard and F. Nejadsattari and Y. Gefen and E. Karimi},
   doi = {10.1038/s42254-019-0071-1},
   issn = {2522-5820},
   journal = {Nat. Rev. Phys.},
   pages = {437},
   title = {{Geometric phase from Aharonov–Bohm to Pancharatnam–Berry and beyond}},
   volume = {1},
   year = {2019}
}

@article{Jisha2021,
   author = {C. P. Jisha and S. Nolte and A. Alberucci},
   doi = {10.1002/lpor.202100003},
    pages = {2100003},
   journal = {Laser \& Photonics Rev.},
   title = {{Geometric Phase in Optics: From Wavefront Manipulation to Waveguiding}},
   volume = {15},
   year = {2021}
}

@article{Marrucci2011JO,
	author = {Marrucci, L. and Karimi, E. and Slussarenko, S. and Piccirillo, B. and Santamato, E. and Nagali, E. and Sciarrino, F.},
	title = {{Spin-to-orbital conversion of the angular momentum of light and its classical and quantumapplications}},
	journal = {J. Opt.},
	volume = {13},
	pages = {064001},
	year = {2011},
	doi = {10.1088/2040-8978/13/6/064001}
}

@Article{Bliokh2015NP,
  author  = {Bliokh, K. Y. and {Rodr{\'i}guez-Fortu{\~n}o}, F. J. and Nori, F. and Zayats, A. V.},
  journal = {Nat. Photonics},
  title   = {Spin-Orbit Interactions of Light},
  year    = {2015},
  pages   = {796},
  volume  = {9},
  doi     = {10.1038/nphoton.2015.201},
}

@article{Berry2009,
   author = {M. V. Berry},
   doi = {10.1088/1464-4258/11/9/094001},
   journal = {J. Opt. A: Pure Appl. Opt.},
   pages = {094001},
   title = {Optical currents},
   volume = {11},
   year = {2009}
}

@article{Bliokh2014,
   author = {K. Y. Bliokh and Y. S. Kivshar and F. Nori},
   doi = {10.1103/PhysRevLett.113.033601},
   journal = {Phys. Rev. Lett.},
   pages = {033601},
   title = {{Magnetoelectric Effects in Local Light-Matter Interactions}},
   volume = {113},
   year = {2014}
}

@article{Hannay1998,
   author = {J. H. Hannay},
   doi = {10.1080/09500349808230892},
   journal = {J. Mod. Opt.},
   pages = {1001},
   title = {The Majorana representation of polarization, and the Berry phase of light},
   volume = {45},
   year = {1998}
}

@article{Padgett1999,
   author = {M. J. Padgett and J. Courtial},
   doi = {10.1364/OL.24.000430},
   journal = {Opt. Lett.},
   pages = {430},
   title = {{Poincaré-sphere equivalent for light beams containing orbital angular momentum}},
   volume = {24},
   year = {1999}
}

@article{vanEnk1993,
   author = {S. J. van Enk},
   doi = {10.1016/0030-4018(93)90472-H},
   journal = {Opt. Commun.},
   month = {9},
   pages = {59},
   title = {Geometric phase, transformations of gaussian light beams and angular momentum transfer},
   volume = {102},
   year = {1993}
}

@article{Rytov1938,
    author = {S. M. Rytov},
    title = {On transition from wave to geometrical optics},
    journal = {Dokl. Akad. Nauk SSSR},
    year = {1938},
    volume = {18},
    pages = {263},
}

@article{Vladimirskiy1941,
    author = {V. V. Vladimirskiy},
    title = {The rotation of a polarization plane for curved light ray},
    journal = {Dokl. Akad. Nauk SSSR},
    year = {1941},
    volume = {31},
    pages = {222},
}

@article{Vernon2025_1,
   author = {A. J. Vernon},
   month = {7},
   title = {Topologies of light in electric-magnetic space},
   year = {2025},
   doi = {10.48550/arXiv.2507.16721},
   journal = {arXiv:2507.16721}
}

@article{Golat2025,
   author = {S. Golat and A. J. Vernon and F. J. Rodríguez-Fortuño},
   doi = {10.1088/1402-4896/ae0662},
   journal = {Phys. Scr.},
   month = {10},
   pages = {105518},
   title = {The electromagnetic symmetry sphere: a framework for energy, momentum, spin and other electromagnetic quantities},
   volume = {100},
   year = {2025}
}

@article{Bliokh2013,
   author = {K. Y. Bliokh and A. Y. Bekshaev and F. Nori},
   doi = {10.1088/1367-2630/15/3/033026},
   journal = {New J. Phys.},
   month = {3},
   pages = {033026},
   title = {Dual electromagnetism: helicity, spin, momentum and angular momentum},
   volume = {15},
   year = {2013}
}

@article{Berry2019,
   author = {M. V. Berry and P. Shukla},
   doi = {10.1088/2040-8986/ab14c4},
   journal = {J. Opt.},
   month = {6},
   pages = {064002},
   title = {{Geometry of 3D monochromatic light: local wavevectors, phases, curl forces, and superoscillations}},
   volume = {21},
   year = {2019}
}

@article{Berry2024,
   author = {M. V. Berry},
   doi = {10.1364/OPN.35.3.000042},
   journal = {Opt. Photonics News},
   month = {3},
   pages = {42},
   title = {{A Geometric-Phase Timeline}},
   volume = {35},
   year = {2024}
}

@article{Cheng2025,
   author = {Y. Cheng and Y.-S. Zeng and W. Xiao and T. Fu and J. Wu and G.-B. Wu and D. P. Tsai and S. Wang},
   title = {{Riemann-Silberstein geometric phase for high-dimensional light manipulation}},
   year = {2025},
   doi = {10.48550/arXiv.2510.09112},
   journal = {arXiv:2510.09112},
}

@article{Dennis2017,
   author = {M. R. Dennis and M. A. Alonso},
   doi = {10.1098/rsta.2015.0441},
   journal = {Philos. Trans. R. Soc. A},
   month = {2},
   pages = {20150441},
   title = {{Swings and roundabouts: optical Poincaré spheres for polarization and Gaussian beams}},
   volume = {375},
   year = {2017}
}

@article{Galvez2003,
   author = {E. J. Galvez and P. R. Crawford and H. I. Sztul and M. J. Pysher and P. J. Haglin and R. E. Williams},
   doi = {10.1103/PhysRevLett.90.203901},
   journal = {Phys. Rev. Lett.},
   month = {5},
   pages = {203901},
   title = {{Geometric Phase Associated with Mode Transformations of Optical Beams Bearing Orbital Angular Momentum}},
   volume = {90},
   year = {2003}
}

@article{Calvo2005,
   author = {G. F. Calvo},
   doi = {10.1364/OL.30.001207},
   journal = {Opt. Lett.},
   month = {5},
   pages = {1207},
   title = {Wigner representation and geometric transformations of optical orbital angular momentum spatial modes},
   volume = {30},
   year = {2005}
}

@article{Green2023,
   author = {D. Green and K. A. Forbes},
   doi = {10.1039/D2NR05426D},
   journal = {Nanoscale},
   pages = {540},
   title = {Optical chirality of vortex beams at the nanoscale},
   volume = {15},
   year = {2023}
}

@article{Vernon2025,
   author = {A. J. Vernon and S. Golat and F. J. Rodríguez-Fortuño},
   doi = {10.1103/9z11-fw8m},
   journal = {Phys. Rev. A},
   volume = {112},
   title = {Electromagnetic symmetry dislocations},
   year = {2025},
   pages = {L021504},
}

@article{Xiao2010RMP,
	author = {Xiao, D. and Chang, M.-C. and Niu, Q.},
	title = {{Berry phase effects on electronic properties}},
	journal = {Rev. Mod. Phys.},
	volume = {82},
	pages = {1959},
	year = {2010},
	doi = {10.1103/RevModPhys.82.1959}
}

@article{Zwanziger1990ARPC,
	author = {Zwanziger, J. W. and Koenig, M. and Pines, A.},
	title = {{Berry's Phase}},
	journal = {Annu. Rev. Phys. Chem.},
    volume = {41},
	pages = {601},
	year = {1990},
	doi = {10.1146/annurev.pc.41.100190.003125}
}

@article{Vinitskii1990PU,
	author = {Vinitskii, S. I. and Derbov, V. L. and Dubovik, Vladimir M. and Markovski, B. L. and Stepanovskii, Y. P.},
	title = {{Topological phases in quantum mechanics and polarization optics}},
	journal = {Sov. Phys. Usp.},
	volume = {33},
	pages = {403},
	year = {1990},
	doi = {10.1070/PU1990v033n06ABEH002598}
}

@article{Bhandari1997PR,
	author = {Bhandari, R.},
	title = {{Polarization of light and topological phases}},
	journal = {Phys. Rep.},
	volume = {281},
	pages = {1},
	year = {1997},
	doi = {10.1016/S0370-1573(96)00029-4}
}

@article{Cisowski2022RMP,
	author = {Cisowski, C. and G{\ifmmode\ddot{o}\else\"{o}\fi}tte, J. B. and Franke-Arnold, S.},
	title = {{Colloquium: Geometric phases of light: Insights from fiber bundle theory}},
	journal = {Rev. Mod. Phys.},
	volume = {94},
	pages = {031001},
	year = {2022},
	doi = {10.1103/RevModPhys.94.031001}
}

@book{Wilczek1989_book,
	author = {Wilczek, F. and Shapere, A.},
	title = {{Geometric Phases in Physics}},
	year = {1989},
	isbn = {978-9971-5-0621-6},
	publisher = {World Scientific Publishing Company},
	address = {Singapore},
	doi = {10.1142/0613}
}

@article{Baggio2017JHEP,
	author = {Baggio, M. and Niarchos, V. and Papadodimas, K.},
	title = {{Aspects of Berry phase in QFT}},
	journal = {{J. High Energy Phys.}},
	volume = {2017},
	pages = {62},
	year = {2017},
	doi = {10.1007/JHEP04(2017)062}
}

@Article{Hasman2005PO,
	author = {Hasman, E. and Biener, G. and Niv, A. and Kleiner, V.},
	title = {{Space-variant polarization manipulation}},
	journal = {{Prog. Opt.}},
	volume = {47},
	pages = {215},
	year = {2005},
	doi = {10.1016/S0079-6638(05)47004-3}
}

@article{Aharonov1987,
   author = {Y. Aharonov and J. Anandan},
   doi = {10.1103/PhysRevLett.58.1593},
   journal = {Phys. Rev. Lett.},
   pages = {1593},
   title = {{Phase change during a cyclic quantum evolution}},
   volume = {58},
   year = {1987}
}

@article{Forbes2021,
   author = {Kayn A Forbes and Garth A Jones},
   doi = {10.1088/2040-8986/ac24bd},
   issn = {2040-8978},
   issue = {11},
   journal = {J. Opt.},
   pages = {115401},
   title = {Measures of helicity and chirality of optical vortex beams},
   volume = {23},
   year = {2021}
}

@article{Bhandari1988PRL,
	author = {Bhandari, R. and Samuel, J.},
	title = {{Observation of topological phase by use of a laser interferometer}},
	journal = {Phys. Rev. Lett.},
	volume = {60},
	pages = {1211},
	year = {1988},
	doi = {10.1103/PhysRevLett.60.1211}
}

@article{Tomita1986,
  title = {{Observation of Berry's Topological Phase by Use of an Optical Fiber}},
  author = {Tomita, A. and Chiao, R. Y.},
  journal = {Phys. Rev. Lett.},
  volume = {57},
  pages = {937},
  year = {1986},
  doi = {10.1103/PhysRevLett.57.937},
}

@article{Ross1984,
  title = {{The rotation of the polarization in low birefringence monomode optical fibres due to geometric effects}},
  author = {Ross, J. N.},
  journal = {Opt. Quantum Electron.},
  volume = {16},
  pages = {455},
  year = {1984},
  doi = {10.1007/BF00619638},
}

@article{BialynickiBirula1996,
title = {{Photon Wave Function}},
journal = {Prog. Opt.},
volume = {36},
pages = {245},
year = {1996},
doi = {https://doi.org/10.1016/S0079-6638(08)70316-0},
url = {https://www.sciencedirect.com/science/article/pii/S0079663808703160},
author = {I. Bialynicki-Birula}
}

@article{Freund2010OC,
	author = {I. Freund},
	journal = {Opt. Commun.},
	pages = {1},
	title = {{Optical M{\"o}bius strips in three-dimensional ellipse fields: I. Lines of circular polarization}},
	doi = {10.1016/j.optcom.2009.09.042},
	volume = {283},
	year = {2010}
}

@article{Bauer2015S,
	author = {T. Bauer and P. Banzer and E. Karimi and S. Orlov and A. Rubano and L. Marrucci and E. Santamato and R. W. Boyd and G. Leuchs},
	journal = {Science},
	pages = {964},
	title = {{Observation of optical polarization M{\"o}bius strips}},
	doi = {10.1126/science.1260635},
	volume = {347},
	year = {2015}
}

@article{Bliokh2021POF,
	author = {Bliokh, K. Y. and Alonso, M. A. and Sugic, D. and Perrin, M. and Nori, F. and Brasselet, E.},
	doi = {10.1063/5.0056333},
	journal = {Phys. Fluids},
	pages = {077122},
	title = {{Polarization singularities and M{\"o}bius strips in sound and water-surface waves}},
	volume = {33},
	year = {2021}
}

@article{Muelas2022PRL,
	author = {Muelas-Hurtado, R. D. and Volke-Sep\'ulveda, K. and Ealo, J. L. and Nori, F. and Alonso, M. A. and Bliokh, K. Y. and Brasselet, E.},
	doi = {10.1103/PhysRevLett.129.204301},
	journal = {Phys. Rev. Lett.},
	pages = {204301},
	title = {Observation of Polarization Singularities and Topological Textures in Sound Waves},
	volume = {129},
	year = {2022}
}

@article{Cameron2012,
	author = {R. P. Cameron and S. M. Barnett and A. M. Yao},
	journal = {New J. Phys.},
	pages = {053050},
	title = {Optical helicity, optical spin and related quantities in electromagnetic theory},
	volume = {14},
	year = {2012},
	doi = {10.1088/1367-2630/14/5/053050}
}

@article{Forbes2025,
   author = {K. A. Forbes},
   month = {8},
   title = {Vortex Light at the Nanoscale: Twists, Spins, and Surprises -- A Review},
   year = {2025},
   journal = {arXiv:2508.09564}
}

@article{Galiffi2022AP,
author = {E. Galiffi and R. Tirole and S. Yin and H. Li and S. Vezzoli and P. A. Huidobro and M. G. Silveirinha and R. Sapienza and A. Al{\`u} and J. B. Pendry},
title = {{Photonics of time-varying media}},
volume = {4},
journal = {Adv. Photonics},
pages = {014002},
year = {2022},
doi = {10.1117/1.AP.4.1.014002}
}

@article{Yessenov2022AOP,
author = {M. Yessenov and L. A. Hall and K. L. Schepler and A. F. Abouraddy},
journal = {Adv. Opt. Photon.},
pages = {455},
title = {Space-time wave packets},
volume = {14},
year = {2022},
doi = {10.1364/AOP.450016}
}

@article{Shen2023JO,
	author = {Y. Shen and Q. Zhan and L. G. Wright and D. N. Christodoulides and F. W. Wise and A. E. Willner and K.-H. Zou and Z. Zhao and M. A. Porras and A. Chong and C. Wan and K. Y. Bliokh and C.-T. Liao and C. Hern{\'a}ndez-Garc{\'\i}a and M. Murnane and M. Yessenov and A. F. Abouraddy and L. J. Wong and M. Go and S. Kumar and C. Guo and S. Fan and N. Papasimakis and N. I. Zheludev and L. Chen and W. Zhu and A. Agrawal and M. Mounaix and N. K. Fontaine and J. Carpenter and S. W. Jolly and C. Dorrer and B. Alonso and I. Lopez-Quintas and M. L{\'o}pez-Ripa and {\'I}. J. Sola and J. Huang and H. Zhang and Z. Ruan and A. H. Dorrah and F. Capasso and A. Forbes},
	doi = {10.1088/2040-8986/ace4dc},
	journal = {J. Opt.},
	pages = {093001},
	title = {Roadmap on spatiotemporal light fields},
	volume = {25},
	year = {2023}
}

\end{document}


\title{Supplementary Information for `Electric-Magnetic Geometric Phase'}

\author{Alex J. Vernon}
\email{alex.vernon@dipc.org}
\affiliation{Donostia International Physics Center (DIPC), Donostia-San Sebasti\'an 20018, Spain}

\author{Konstantin Y. Bliokh}
\email{konstantin.bliokh@dipc.org}
\affiliation{Donostia International Physics Center (DIPC), Donostia-San Sebasti\'an 20018, Spain}
\affiliation{IKERBASQUE, Basque Foundation for Science, Bilbao 48009, Spain}
\affiliation{Centre of Excellence ENSEMBLE3 Sp.~z o.o., 01-919 Warsaw, Poland}

\maketitle
This supplementary document has three main parts.
The first, \ref{sec1}, explicitly shows why the EM geometric phase must be zero in paraxial waves as is stated in the main text after equation 9.
The second, \ref{sec2}, shows how when in a nonparaxial field both electric and magnetic fields have the same, constant polarisation ellipse, the EM geometric phase can be directly associated with a closed path on a Poincar\'e-like sphere (the sphere having been proposed in \cite{Golat2025}), justifying equation 11 of the main text.
The final section, \ref{sec3}, gives full derivations for equation 18 of the main text and further background to Fig.~2, demonstrating how the EM geometric phase can arise in transformations to the longitudinal components of focussed beams. 

Before proceeding it will be useful to recap and reprint key expressions from the main text here for later reference.
We described electromagnetic fields using the $\mathbb{C}^2\otimes\mathbb{C}^3\otimes\mathbb{L}^2$ vector $\Bisp$, a function of some parameters $\bm{\rho}$ which can be expressed using a normalised state vector $\bisp$,
\begin{equation}\label{bispinor}
\begin{split}
    \Bisp(\bm{\rho})&=\frac12\begin{pmatrix}\sqrt{\varepsilon_0}\mathbf{E}(\bm{\rho})\\\sqrt{\mu_0}\mathbf{H}(\bm{\rho})\end{pmatrix}=\sqrt{W(\bm{\rho})}\ee^{\ii\alpha(\bm{\rho})}\underbrace{\begin{pmatrix}
        \cos{\frac{\theta(\bm{\rho})}{2}}\ee^{\ii\chi_\text{e}(\bm{\rho})}\unit{e}\\\sin{\frac{\theta(\bm{\rho})}{2}}\ee^{\ii\chi_\text{m}(\bm{\rho})}\unit{h}
    \end{pmatrix}}_{\bisp(\bm{\rho})},
\end{split}
\end{equation}
where the phase angle $\alpha$ was chosen to be,
\begin{equation}\label{alpha}
\alpha=\frac12\Arg\left(\Bisp^\intercal\cdot\Bisp\right)=\frac12\Arg\left(\frac14\varepsilon_0\mathbf{E}\cdot\mathbf{E}+\frac14\mu_0\mathbf{H}\cdot\mathbf{H}\right).
\end{equation}
This and all other phase angles in this work are defined modulo $2\pi$.
This choice is not unique but is useful because, from \eqref{bispinor}, it fixes $\Arg\left(\bisp(\bm{\rho})^\intercal\cdot\bisp(\bm{\rho})\right)=0$ for all $\bm{\rho}$.
If the field $\Bisp$ evolves over a curve $\bm{\rho}=\bm{\rho}_C$ such that the field at the ends of $C$, $\Bisp_\text{in}=\Bisp(\bm{\rho}_\text{in})$ and $\Bisp_\text{fin}=\Bisp(\bm{\rho}_\text{fin})$, only differs by a global phase,
i.e.,
\begin{equation}\label{in_fin_relation}
    \Bisp_\text{fin}=\ee^{\ii\Phi_\text{glo}}\Bisp_\text{in},
\end{equation}
from which $\Phi_\text{glo}$ can be found using,
\begin{equation}\label{phi_glo}
    \Phi_\text{glo}=\Arg{\left(\Bisp^\dagger_\text{in}\cdot\Bisp_\text{fin}\right)}=\Delta\alpha+\Arg{\left(\bisp^\dagger_\text{in}\cdot\bisp_\text{fin}\right)},
\end{equation}
then it follows that,
\begin{equation}
\begin{split}
    \Arg\left(\bisp_\text{in}^\intercal\cdot\bisp_\text{in}\right)&=2(\Delta\alpha-\Phi_\text{glo})+\Arg\left(\bisp^\intercal_\text{fin}\cdot\bisp_\text{fin}\right)\\
    &\to2(\Delta\alpha-\Phi_\text{glo})=0.
\end{split}
\end{equation}
Because $2(\Delta\alpha-\Phi_\text{glo})=0$ is true modulo $2\pi$, we can write
\begin{equation}
    \Phi_\text{glo}=\Delta\alpha+N\pi
\end{equation}
where $N=0,1$ is an integer, therefore accounting for the second term in \eqref{phi_glo}, that is,
\begin{equation}\label{Npi}
    N\pi=\Arg\left(\bisp^\dagger_\text{in}\cdot\bisp_\text{fin}\right).
\end{equation}
In \eqref{bispinor}, the unit polarisation field vectors $\unit{e}$ and $\unit{h}$ are,
\begin{equation}
\begin{split}
\unit{e}&=\ee^{-\ii(\alpha+\chi_\text{e})}\mathbf{E}/|\mathbf{E}|,\\
\unit{h}&=\ee^{-\ii(\alpha+\chi_\text{m})}\mathbf{H}/|\mathbf{H}|.
\end{split}
\end{equation}
The phase angles $\chi_\text{e}$ and $\chi_\text{m}$ are defined as
\begin{align}
    \chi_\text{e}&=\frac12\Arg\left(\mathbf{E}\cdot\mathbf{E}\right)-\alpha,\label{chie}\\
    \chi_\text{m}&=\frac12\Arg\left(\mathbf{H}\cdot\mathbf{H}\right)-\alpha,\label{chim}
\end{align}
in order to enforce $\Arg(\unit{e}\cdot\unit{e})=\Arg(\unit{h}\cdot\unit{h})=0$.
Later on in the main text we arrived at the expression of geometric phase $\Phi_\text{g}$,
\begin{equation}\label{phi_gsplit}
    \Phi_\text{g}=\Phi_\text{gI}+\Phi_\text{gII}+N\pi
\end{equation}
The contributing terms in \eqref{phi_gsplit} are,
\begin{equation}\label{phi_g1}
\begin{split}
        \Phi_\text{gI}=\ii\int_C\left[\cos^2{\frac{\theta}{2}}\unit{e}^*\cdot(\nabla_\mathbf{r})\unit{e}\cdot+\sin^2{\frac{\theta}{2}}\unit{h}^*\cdot(\nabla_\mathbf{r})\unit{h}\right]\cdot d\mathbf{r},
\end{split}
\end{equation}
and
\begin{equation}\label{phi_g2}
    \Phi_\text{gII}=-\int_C\left[\cos^2{\frac{\theta}{2}}d\chi_\text{e}+\sin^2{\frac{\theta}{2}}d\chi_\text{m}\right],
\end{equation}
which is the topic of this work, the EM geometric phase.
We emphasise again that expressions of phase angles, like \eqref{phi_gsplit}, \eqref{phi_g1} and \eqref{phi_g2}, are taken modulo $2\pi$.

\section{Geometric phase in paraxial light}\label{sec1}
In the main text we said that in paraxial light, the electric and magnetic fields are constrained to a degree that the EM geometric phase $\Phi_\text{gII}$ must be zero.
This results from the close relation between electric and magnetic polarisation in paraxial waves: the two fields are related by $\eta\mathbf{H}=\unit{k}\times\mathbf{E}$, where $\eta=\sqrt{\mu_0/\varepsilon_0}$ and $\unit{k}$ is the unit vector in the direction of propagation ($\unit{k}\cdot\unit{k}=1$), to which both $\mathbf{E}$ and $\mathbf{H}$ are transverse ($\mathbf{E}\cdot\unit{k}=\mathbf{H}\cdot\unit{k}=0$).
It follows that
\begin{equation}
\begin{split}
    \Arg(\mathbf{H}\cdot\mathbf{H})&=\Arg\left([\unit{k}\times\mathbf{E}]\cdot[\unit{k}\times\mathbf{E}]\right)\\
    &=\Arg\left([\unit{k}\cdot\unit{k}][\mathbf{E}\cdot\mathbf{E}]-[\unit{k}\cdot\unit{E}]^2\right)\\
    &=\Arg(\mathbf{E}\cdot\mathbf{E}).
\end{split}
\end{equation}
This means, from \eqref{alpha}, $\alpha=\Arg(\mathbf{E}\cdot\mathbf{E})/2=\Arg(\mathbf{H}\cdot\mathbf{H})/2$, so it must be that in equations \eqref{chie} and \eqref{chim}, $\chi_\text{e}=\chi_\text{m}=0$, and subsequently $\Phi_\text{gII}=0$ \eqref{phi_g2}.
Geometric phase in paraxial light can therefore only be due to $\Phi_\text{gI}$, ie., $\Phi_\text{g}=\Phi_\text{gI}+N\pi$, which emerges when the electric and magnetic polarisation states evolve cyclically---the well known mechanism for the Pancharatnam-Berry phase (this is assuming that the propagation direction of a paraxial wave stays in the same plane while its polarisation evolves so that the spin-redirection phase is zero).
Note that $\Phi_\text{g}=\Phi_\text{PB}$ in paraxial waves is invariant to changes of the EM-space basis of $\Bisp$.

\section{EM geometric phase and the Poincar\'e-like sphere}\label{sec2}
We said in the manuscript that when an electromagnetic field $\Bisp$ evolves cylclically while the polarisation states of the electric and magnetic fields are the same and constant (meaning $\unit{e}=\unit{h}=\text{constant}$), then any geometric phase incurred due to the evolution can be associated unambiguously with our proposed EM geometric phase, that is,
\begin{equation}\label{em_phase_condition}
    \Phi_\text{g}=\Phi_\text{gII}+N\pi\quad[\text{when }\unit{e}=\unit{h}=\text{constant}].
\end{equation}
Our purpose in this section is to show how this corresponds to the solid angle swept out by a Stokes-like vector $\mathbf{W}$ whose components $(W_1,W_2,W_3)$ are given by,
\begin{equation}\label{W_i}
    W_i=\frac{1}{W}\Bisp^\dagger\cdot(\hat{\sigma}_i)\Bisp
\end{equation}
where $\hat{\sigma}_i$ is the $i^\text{th}$ Pauli matrix.
Recall from the main text equation 10, where $\mathbf{W}$ is given explicitly as,
\begin{equation}
    \mathbf{W}=\frac{1}{4W}\begin{pmatrix}
        2\Re\{\mathbf{E}^*\cdot\mathbf{H}\}/\omega c\\
        2\Im\{\mathbf{E}^*\cdot\mathbf{H}\}/\omega c\\
        \varepsilon_0|\mathbf{E}|^2-\mu_0|\mathbf{H}|^2
    \end{pmatrix},
\end{equation}
where $W=(\varepsilon_0|\mathbf{E}|^2+\mu_0|\mathbf{H}|^2)/4$.
Note that if the basis of $\Bisp$ is changed in \eqref{W_i} then the order of the components of $\mathbf{W}$ changes (just like the Stokes vector for 2D electric polarisation).
Assuming $\unit{e}^*\cdot\unit{h}=1$, it follows that,
\begin{equation}
\begin{split}
    \mathbf{E}^*\cdot\mathbf{H}&=|\mathbf{E}||\mathbf{H}|\ee^{-\ii(\alpha+\chi_\text{e})}\ee^{\ii(\alpha+\chi_\text{m})}\unit{e}^*\cdot\unit{h}\\
    &=|\mathbf{E}||\mathbf{H}|\ee^{\ii(\chi_\text{m}-\chi_\text{e})}
\end{split}
\end{equation}
meaning we can define the azimuth angle $\phi=\arctan(\Im\{\mathbf{E}^*\cdot\mathbf{H}\}/\Re\{\mathbf{E}^*\cdot\mathbf{H}\})=\chi_\text{m}-\chi_\text{e}$.
Given the elevation angle
\begin{equation}
    \theta=2\arctan\left(\frac{\sqrt{\mu_0}|\mathbf{H}|}{\sqrt{\varepsilon_0}|\mathbf{E}|}\right)=\arccos\left(\frac{\varepsilon_0|\mathbf{E}|^2-\mu_0|\mathbf{H}|^2}{4W}\right),
\end{equation}
$\mathbf{W}$ can be written as
\begin{equation}
    \mathbf{W}=\begin{pmatrix}
        \sin{\theta}\cos{\phi}\\\sin{\theta}\sin{\phi}\\\cos\theta\end{pmatrix}.
\end{equation}
If $\theta$ and $\phi$ vary such that the tip of $\mathbf{W}$ sweeps out a closed path, the enclosed solid angle is,
\begin{equation}
    \Omega_\mathbf{W}=\oint(1-\cos\theta)d\phi.
\end{equation}
Returning now to \eqref{em_phase_condition}, we can use trigonometric identities for $\cos^2\theta/2$ and $\sin^2\theta/2$ in \eqref{phi_g2} to write,
\begin{equation}\label{phi_g_alt}
\begin{split}
    \Phi_\text{g}&=\left[-\frac12\int_C(d\chi_\text{e}+d\chi_\text{m})+\frac12\int_C\cos\theta(d\chi_\text{m}-d\chi_\text{e})\right]\bmod2\pi+N\pi.
\end{split}
\end{equation}
We have explicitly indicated above that $\Phi_\text{gII}$ \eqref{phi_g2} is taken modulo $2\pi$, which will be helpful in the following calculations.
Using \eqref{Npi} and the fact that $\theta(\bm{\rho}_\text{in})=\theta(\bm{\rho}_\text{fin})$ (due to Eq.~\ref{in_fin_relation}), we have,
\begin{equation}\label{phi_glo_bisp}
\begin{split}
    \Arg\left(\bisp^\dagger_\text{in}\cdot\bisp_\text{fin}\right)&=\Arg\left(\cos^2{\frac{\theta_\text{in}}{2}}\ee^{\ii(\chi_\text{e}(\bm{\rho}_\text{fin})-\chi_\text{e}(\bm{\rho}_\text{in}))}+\sin^2{\frac{\theta_\text{in}}{2}}\ee^{\ii(\chi_\text{m}(\bm{\rho}_\text{fin})-\chi_\text{m}(\bm{\rho}_\text{in}))}\right)=N\pi.
    \end{split}
\end{equation}
Also due to Eq.~\ref{in_fin_relation} is the fact that $\chi_\text{m}(\bm{\rho}_\text{fin})-\chi_\text{e}(\bm{\rho}_\text{fin})=\chi_\text{m}(\bm{\rho}_\text{in})-\chi_\text{e}(\bm{\rho}_\text{in})$.
From \eqref{phi_glo_bisp} it then follows that
\begin{equation}
\begin{split}
    \chi_\text{e}(\bm{\rho}_\text{fin})-\chi_\text{e}(\bm{\rho}_\text{in})&=\int_Cd\chi_\text{e}\bmod2\pi=N\pi\\
    \chi_\text{m}(\bm{\rho}_\text{fin})-\chi_\text{m}(\bm{\rho}_\text{in})&=\int_Cd\chi_\text{m}\bmod2\pi=N\pi.
    \end{split}
\end{equation}
Now writing
\begin{equation}
     \int_Cd\chi_\text{e}=N\pi+2\pi n_\text{e};\quad\int_Cd\chi_\text{m}=N\pi+2\pi n_\text{m},
\end{equation}
where $n_\text{e},n_\text{m}$ are integers, we find that
\begin{equation}
    \frac{1}{2}\int_C(d\chi_\text{e}+d\chi_\text{m})\bmod2\pi=N\pi+[(n_\text{e}+n_\text{m})\bmod2]\pi,
\end{equation}
while,
\begin{equation}
    \frac{1}{2}\int_C(d\chi_\text{m}-d\chi_\text{e})\bmod2\pi=[(n_\text{m}-n_\text{e})\bmod2]\pi.
\end{equation}
Since the sum $n_\text{e}+n_\text{m}$ and difference $n_\text{m}-n_\text{e}$ of two integers always has the same parity, we can conclude
\begin{equation}
    -\left[\frac12\int_C(d\chi_\text{e}+d\chi_\text{m})\right]\bmod2\pi+N\pi\equiv-\frac{1}{2}\int_C(d\chi_\text{m}-d\chi_\text{e})\bmod2\pi.
\end{equation}
Substituting this in \eqref{phi_g_alt}, we have,
\begin{equation}
\begin{split}
    \Phi_\text{g}&=-\frac{1}{2}\int_C(d\chi_\text{m}-d\chi_\text{e})+\frac12\int_C\cos\theta(d\chi_\text{m}-d\chi_\text{e})\mod2\pi\\
    &=-\frac12\int_C(1-\cos\theta)d\phi\mod2\pi\\
    &=-\frac12\Omega_\mathbf{W}\mod2\pi.
\end{split}
\end{equation}
We have now arrived at what was presented in equation 11 of the main text.

\section{Modal and EM geometric phases}\label{sec3}

We said that the transverse field of a general, first-order paraxial beam can be represented by,
\begin{equation}\label{bisp_f}
\begin{split}
    \Bisp^\perp_f(\mathbf{r},t)&=\ee^{-\ii\omega_0 t}\Bisp_0^\perp\underbrace{\left[a(t)\text{LG}^{-1}_0(\mathbf{r})+b(t)\text{LG}^{1}_0(\mathbf{r})\right]}_{f(\mathbf{r},t)},
\end{split}
\end{equation}
in which the scalar field $f(\mathbf{r},t)$ carries the spatial structure of the beam which may be made to evolve over time $t$, $\Bisp_0^\perp$ is a constant vector determining the beam's transverse polarisation, and the super- and sub-scripts of $\text{LG}^l_p$ are topological charge and radial order respectively.
It will be useful to work in a basis of Riemann-Silberstein vectors going forward, i.e., $\Bisp_0^\perp=(\mathbf{F}_\text{R0}^\perp,\mathbf{F}_\text{L0}^\perp)/2$.
If the transverse electric and magnetic fields of the beam are given by,
\begin{equation}
    \mathbf{E}_0^\perp=E_{x0}\unit{x}+E_{y0}\unit{y};\quad\mathbf{H}_0^\perp=\frac1\eta\left(-E_{y0}\unit{x}+E_{x0}\unit{y}\right),
\end{equation}
where $E_{x0}$ and $E_{y0}$ are complex, then the Riemann-Silberstein vectors $\mathbf{F}_{\text{R}0}^\perp=(\sqrt{\varepsilon_0}\mathbf{E}^\perp_0+\ii\sqrt{\mu_0}\mathbf{H}^\perp_0)/\sqrt{2}$ and $\mathbf{F}_{\text{L}0}^\perp=(\sqrt{\varepsilon_0}\mathbf{E}^\perp_0-\ii\sqrt{\mu_0}\mathbf{H}^\perp_0)/\sqrt{2}$ are,
\begin{equation}
    \mathbf{F}_{\text{R}0}^\perp=\sqrt{\varepsilon_0}(E_{x0}-\ii E_{y0})\left(\frac{\unit{x}+\ii\unit{y}}{\sqrt{2}}\right);\quad\mathbf{F}_{\text{L}0}^\perp=\sqrt{\varepsilon_0}(E_{x0}+\ii E_{y0})\left(\frac{\unit{x}-\ii\unit{y}}{\sqrt{2}}\right),
\end{equation}
which are circularly polarised in the transverse plane with opposite senses.
This means $\Bisp_0^\perp$ can be expressed as,
\begin{equation}\label{bisp0}
    \Bisp_0^\perp=\frac12\begin{pmatrix}
        \mathbf{F}_\text{R0}^\perp\\\mathbf{F}_\text{L0}^\perp
    \end{pmatrix}=\frac{1}{2}\begin{pmatrix}
        A_\text{R}^\perp\left[\frac{\unit{x}+\ii\unit{y}}{\sqrt{2}}\right]\\A_\text{L}^\perp\left[\frac{\unit{x}-\ii\unit{y}}{\sqrt{2}}\right]
    \end{pmatrix}.
\end{equation}
where $A_\text{R/L}^\perp=\sqrt{\varepsilon_0}(E_{x0}\mp\ii E_{y0})$ are the complex amplitudes of right/left circular components.

In the main text we showed that after a first-order approximation of longitudinal components, the geometric phase which would arise in a paraxial beam due to transformations over the modal Poincar\'e sphere manifests as an EM geometric phase in the centre of a focussed beam.
We approximated longitudinal ($z$) components of a general focussed beam as
\begin{equation}\label{psi_z}
\begin{split}
    \Bisp_f^\parallel&\approx\unit{z}\frac{\ii}{k}\nabla_\perp\cdot\Bisp_f^\perp=\unit{z}\frac{\ii}{k}\ee^{-\ii\omega_0 t}\Bisp_0\cdot \left[a(t)\nabla_\perp \text{LG}^{-1}_0+b(t)\nabla_\perp \text{LG}^1_0\right],
\end{split}
\end{equation}
which relies on the calculation of the transverse gradient $\nabla_\perp$ of a scalar Laguerre-Gauss function.
In cylindrical co-ordinates, $\nabla_\perp$ is,
\begin{equation}
    \nabla_\perp=\boldsymbol{\hat{\rho}}\frac{\partial}{\partial\rho}+\boldsymbol{\hat{\phi}}\frac1\rho\frac{\partial}{\partial\phi}.
\end{equation}
Above $\hat{\boldsymbol{\rho}}$ is the radial unit vector in cylindrical co-ordinates, not to be confused with the general parameter space vector $\boldsymbol{\rho}$ used throughout the main text.
In its real-space representation in cylindrical co-ordinates, a scalar Laguerre-Gauss function $\text{LG}^l_p$ of general order $p$ and topological charge $l$ has the form \cite{Green2023},
\begin{equation}\label{LG_r}
    \text{LG}^l_p(\rho,\phi,z)=\sqrt{\frac{2p!}{\pi w_0^2[p+|l|!]}}\frac{w_0}{w(z)}\left[\frac{\sqrt{2}\rho}{w(z)}\right]^{|l|}L_p^{|l|}(a)\ee^{\frac{-\rho^2}{w^2(z)}}\ee^{\ii\left[kz+l\phi+\frac{k\rho^2}{2R(z)}-[2p+|l|+1]\zeta(z)\right]}.
\end{equation}
Here, $w_0$ is the beam waist in the focal plane, $w(z)=w_0\sqrt{1+z^2/z_\text{R}^2}$ for which the Rayleigh range is $z_\text{R}=kw_0^2/2$, the wavefront curvature is $R(z)=z-z_\text{R}^2/z$, and $L_p^{|l|}(a)$ is the generalised Laguerre function with argument $a=2\rho^2/w^2(z)$.
Meanwhile $\zeta(z)=\arctan z/z_\text{R}$ is the Gouy phase.

When dealing with beams with mode order 1, i.e., with zero radial index, $p=0$, and $l=\pm1$, \eqref{LG_r} is greatly simplified in the focal plane ($z=0$) to
\begin{equation}\label{LG_p0_l1}
    \text{LG}^{l=\pm1}_{0}(\rho,\phi,z=0)=\frac{2\rho}{\sqrt{\pi}w_0^2}\ee^{\frac{-\rho^2}{w_0^2}}\ee^{\ii l\phi}.
\end{equation}
Then the transverse gradient of \eqref{LG_p0_l1} is,
\begin{equation}
\begin{split}
    \nabla_\perp\text{LG}^{l=\pm1}_0=\hat{\boldsymbol{\rho}}\frac{2}{\sqrt{\pi}w_0^2}\ee^{\ii l\phi}\ee^{\frac{-\rho^2}{w_0^2}}\left[1-\frac{2\rho^2}{w_0^2}\right]+\hat{\boldsymbol{\phi}}\frac{\ii2l}{\sqrt{\pi}w_0^2}\ee^{\ii l\phi}\ee^{\frac{-\rho^2}{w_0^2}}.
\end{split}
\end{equation}
Evaluating the above in the beam's vortex core $\rho=0$, and converting to cartesian components, we have,
\begin{equation}
    \nabla_\perp\text{LG}^{l=\pm1}_0=B\ee^{\ii l \phi}[(\cos\phi-\ii l\sin\phi)\unit{x}+(\ii l\cos\phi+\sin\phi)\unit{y}],
\end{equation}
where $B=2/(\sqrt{\pi}w_0^2)$ is a constant.
Taking the dot product between $\Bisp_0^\perp$ \eqref{bisp0} and $\nabla_\perp\text{LG}^{l=\pm1}_0$ for each topological charge yields,
\begin{equation}\label{cases}
    \Bisp_0^\perp\cdot(\nabla_\perp\text{LG}^{l=\pm1}_0)=\frac{B}{2\sqrt{2}}\begin{pmatrix}
        A_\text{R}^\perp\ee^{\ii(l+1)\phi}(1-l)\\A_\text{L}^\perp\ee^{\ii(l-1)\phi}(1+l)
    \end{pmatrix}=\begin{cases}
    \frac{B}{\sqrt{2}}\begin{pmatrix}
        0\\A_\text{L}^\perp
    \end{pmatrix}\text{ if $l=1$},\\
    \frac{B}{\sqrt{2}}\begin{pmatrix}
        A_\text{R}^\perp\\0
    \end{pmatrix}\text{ if $l=-1$}.
    \end{cases}
\end{equation}
Note that the difference in energy carried by the Riemann-Silberstein vectors $|\mathbf{F}_\text{R}|^2-|\mathbf{F}_\text{L}|^2$ is associated with optical chirality---so in the above equation, a focussed $l=1$ vortex beam in general carries the helicity per photon of $-1$ in the centre of its focus due to longitudinal components (and vice versa for the $l=-1$ beam).
This matches previous studies of optical chirality in vortex beams \cite{Green2023,Forbes2021}.
Substituting \eqref{cases} in \eqref{psi_z}, we have (again, in the centre of the focus),
\begin{equation}\label{longitudinal_field}
    \Bisp_f^\parallel=\frac{\ii B\ee^{-\ii\omega_0 t}}{k\sqrt{2}}\begin{pmatrix}
        A_\text{R}^\perp a(t)\unit{z}\\A_\text{L}^\perp b(t)\unit{z}
    \end{pmatrix}.
\end{equation}
Having produced equation 18 in the main text (corresponding to \eqref{longitudinal_field}), we demonstrated in Fig.~3 of the main text how cyclic modulation over time to the mode $f$, by choosing periodic functions $a(t)$ and $b(t)$, can result in the EM geometric phase.
These were taken to be,
\begin{equation}
    a(t)=\cos\frac{\theta_0}{2}\exp\left[\ii\frac{2\pi t}{T}\sin^2\frac{\theta_0}{2}\right];\quad b(t)=\sin\frac{\theta_0}{2}\exp\left[-\ii\frac{2\pi t}{T}\cos^2\frac{\theta_0}{2}+\ii\chi_0\right],
\end{equation}
where $\theta_0$ is a constant, chosen in Fig.~3 of the main text to be $\pi/6$, and $\chi_0=\pi/2$.
The input paraxial beam was also assumed to be linearly polarised (in the electric and magnetic fields) such that $|A_\text{R}^\perp|=|A_\text{L}^\perp|$.
Note that if the beam is circularly polarised (i.e., one of either $A_\text{R}^\perp$ or $A_\text{L}^\perp$ is zero) then one component of \eqref{longitudinal_field} is always zero and it is not possible for the EM geometric phase to manifest.

\bibliography{bibliography}